\documentclass[aps, prb,  reprint,  superscriptaddress,  longbibliography,  showkeys,  citeautoscript,]{revtex4-2}
\usepackage{silence}
\usepackage[intlimits]{amsmath}
\usepackage{amssymb}
\usepackage{amsthm}
\usepackage{booktabs}
\usepackage{braket}
\usepackage{color}
\usepackage{graphicx}

\usepackage[colorlinks=true,
            linkcolor=black,
            citecolor=black,
            filecolor=black,
            urlcolor=black
            ]{hyperref}
\usepackage[utf8]{inputenc}
\usepackage[version=4]{mhchem}
\usepackage{multirow}
\usepackage{soul}
\usepackage{tabularx}
\usepackage{units}
\usepackage[dvipsnames]{xcolor}

\usepackage{dcolumn}
\usepackage{booktabs}
\usepackage{miller}
\usepackage{sidecap} % For Figure 1 side caption
\newcommand{\rev}{\textcolor{black}}

\makeatletter
\def\@hangfrom@section#1#2#3{\@hangfrom{#1#2}#3}%\MakeTextUppercase{#3}}%
\def\@hangfroms@section#1#2{#1#2}%\MakeTextUppercase{#2}}%
\makeatother

\setcitestyle{super}

\newcommand{\dtu}{
    Department of Electrical and Photonics Engineering, Technical University of Denmark,
    2800 Kgs. Lyngby, Denmark
}

\makeatletter
\newcommand\emailx[1]{%
\move@AF%
\def\@affil{{\normalfont\,#1\strut}{}}%
}%

\begin{document}
%TC:ignore
\title{Integrated on-chip quantum light sources on a van der Waals platform}

\author{Pietro Metuh}
\author{Paweł Wyborski}
\author{Athanasios Paralikis}
\affiliation{\dtu}
\author{Frederik Schröder}
\author{Nicolas Stenger}
\affiliation{\dtu}
\affiliation{NanoPhoton – Center for Nanophotonics, Technical University of Denmark, 2800 Kgs. Lyngby, Denmark}
\author{Niels Gregersen}
\author{Battulga Munkhbat}
\email{bamunk@dtu.dk}
\affiliation{\dtu}

\begin{abstract}  

Scalable photonic quantum information technologies require a platform combining quantum light sources, waveguides, and detectors on a single chip. Here, we introduce a van der Waals platform comprising strain-engineered bilayer WSe$_2$ quantum emitters, integrated on multimode WS$_2$ waveguides with grating couplers, enabling efficient on-chip quantum light sources. The emitters exhibit bright, highly polarised emission that couples efficiently into WS$_2$ waveguides. Under \rev{discrete-state} excitation, we observe high-purity, waveguide-coupled single-photon emission, measured using both \rev{single-port and cross-correlated, multi-port photon streams, yielding $g^{(2)}(0) = 0.003^{+0.030}_{-0.003}$ and $g^{(2)}(0) = 0.076\pm0.023$, respectively}. For a single output, the out-coupled single-photon count rate at the first lens reaches approximately \rev{357}~kHz under continuous-wave \rev{discrete-state} excitation, corresponding to an estimated \rev{minimum} waveguide-coupled rate of \rev{1.68}~MHz. These results demonstrate an efficient, integrated single-photon source and establish a pathway toward scalable photonic quantum information processing centred around nanoengineered van der Waals materials.

\end{abstract}
\maketitle
%TC:endignore

A vision for scalable photonic quantum information processing is the realisation of a quantum photonic circuit by integrating all the building blocks on a single chip \cite{obrienPhotonicQuantumTechnologies2009, zhongQuantumComputationalAdvantage2020, wangScalableBosonSampling2018}. Currently, however, state-of-the-art implementations rely on disparate material systems \cite{pelucchiPotentialGlobalOutlook2021}: silicon \cite{paesaniNearidealSpontaneousPhoton2020, silverstoneSiliconQuantumPhotonics2016} or silicon nitride \cite{wangDeterministicPhotonSource2023} for photonic circuits, III–V semiconductors for deterministic quantum emitters \cite{senellartHighperformanceSemiconductorQuantumdot2017, dingHighefficiencySinglephotonSource2025}, and amorphous superconductors such as NbTiN or MoSi for single-photon detectors \cite{esmaeilzadehSuperconductingNanowireSinglephoton2021}. Integrating these components is possible in some circumstances \cite{schwartzFullyOnChipSinglePhoton2018, reithmaierOnchipTimeResolved2013}, but it requires complex hybrid fabrication with stringent thermal and chemical compatibility, which has become a major bottleneck for scaling up photonic quantum technologies \cite{pelucchiPotentialGlobalOutlook2021}. It is therefore desirable to identify a single material platform that enables monolithic \rev{or hybrid} integration of all essential quantum photonic components \cite{turunenQuantumPhotonicsLayered2022}.

Transition metal dichalcogenides (TMDs) have emerged as a powerful material platform in nano- and quantum photonics owing to their versatility afforded by van der Waals (vdW) assembly \cite{zotevNanophotonicsMultilayerVan2025, montblanchLayeredMaterialsPlatform2023, azimiPhotonicsFlatlandChallenges2025}. Their stackable layered structure \cite{castellanos-gomezDeterministicTransferTwodimensional2014} enables combination of distinct TMDs to engineer new optical and electronic functionalities and straightforward integration with existing photonic platforms. Several mono- and bilayer TMD semiconductors, such as WSe$_2$ \cite{koperskiSinglePhotonEmitters2015, barboneChargetuneableBiexcitonComplexes2018, partoDefectStrainEngineering2021, drawerMonolayerBasedSinglePhotonSource2023, paralikisTailoringPolarizationWSe22024} and MoTe$_2$ \cite{zhaoSitecontrolledTelecomwavelengthSinglephoton2021, wyborskiTriggeredGenerationIndistinguishable2025}, can host localised quantum emitters created by strain or defect engineering \cite{linhartLocalizedIntervalleyDefect2019}, producing single photons between the visible and telecom range \cite{michaelisdevasconcellosSinglePhotonEmittersLayered2022}. In parallel, high nonlinear coefficients in TMDs have been exploited for enhancing nonlinearities in established material platforms \cite{wangEnhancingSi3N4Waveguide2021}.
Several multilayer TMDs have also high refractive indices and low absorption over a broad spectral window, supporting low-loss, tightly confined optical modes suitable for waveguides and photonic cavities \cite{lingAllVanWaals2021, vyshnevyyVanWaalsMaterials2023, leeUltrathinWS2Polariton2023, mooreVanWaalsWaveguide2025}, and facilitated by their atomically smooth surfaces \cite{munkhbatNanostructuredTransitionMetal2023}. 
Beyond light sources and passive photonics, superconducting TMDs such as NbSe$_2$ have recently been explored as ultrathin single-photon detectors \cite{metuhSinglePhotonDetectionSuperconducting2025, zuglianiSinglephotonDetectionFewlayer2026, wangEmergingSinglePhotonDetectors2022}. Together, these complementary features render TMDs a versatile platform for engineering emitters, circuitry, and detectors.

Despite substantial progress in developing individual vdW photonic components \cite{errando-herranzResonanceFluorescenceWaveguideCoupled2021, vyshnevyyVanWaalsMaterials2023, blauthCouplingSinglePhotons2018, huangOnchipIntegratedLight2025, gerardTopdownIntegrationHBN2023}, a complete vdW integrated circuit has not yet been realised. 
\rev{One of the main challenges encountered when integrating van der Waals materials is the emitter-waveguide coupling efficiency $\beta$} \cite{michaelisdevasconcellosSinglePhotonEmittersLayered2022}.
Monolithic integration \rev{maximises $\beta$ but is restricted to materials such as hexagonal boron nitride (hBN), in which quantum emitters form without applied strain} \cite{kimPhotonicCrystalCavities2018, khelifaWSe2LightEmittingDevice2023, liIntegrationHBNQuantum2021, gerardTopdownIntegrationHBN2023}.
\rev{In TMDs, where emitters are strain-induced, the emitter host must be transferred onto the circuit, and the resulting mono- or bilayer is far too thin to reshape the guided modes it couples into} \cite{kimIntegratedChipPlatform2019, yamashitaSpatiallyDeterministicIntegration2025}.
\rev{Earlier demonstrations have targeted coupling to the fundamental mode, for which the field overlap with dipole emitters on the surface is weak \cite{michaelisdevasconcellosSinglePhotonEmittersLayered2022}. The consequence is visible in the reported figures of merit: across prior waveguide-coupled vdW emitters, the second-order correlation at the waveguide output has often remained near or above the $g^{(2)}(0) = 0.5$ threshold \cite{liIntegrationHBNQuantum2021, peyskensIntegrationSinglePhoton2019}, and waveguide-coupled single-photon rates have generally remained well below the MHz threshold \cite{errando-herranzResonanceFluorescenceWaveguideCoupled2021, peyskensIntegrationSinglePhoton2019, liIntegrationHBNQuantum2021}. In addition, while two reports of monolithically integrated hBN emitters have been successfully demonstrated \cite{liIntegrationHBNQuantum2021, gerardTopdownIntegrationHBN2023}, no TMD emitter has been integrated onto a vdW waveguide to the best of our knowledge.}

\begin{SCfigure*}[0.5]
    \includegraphics[width=1.5\linewidth]{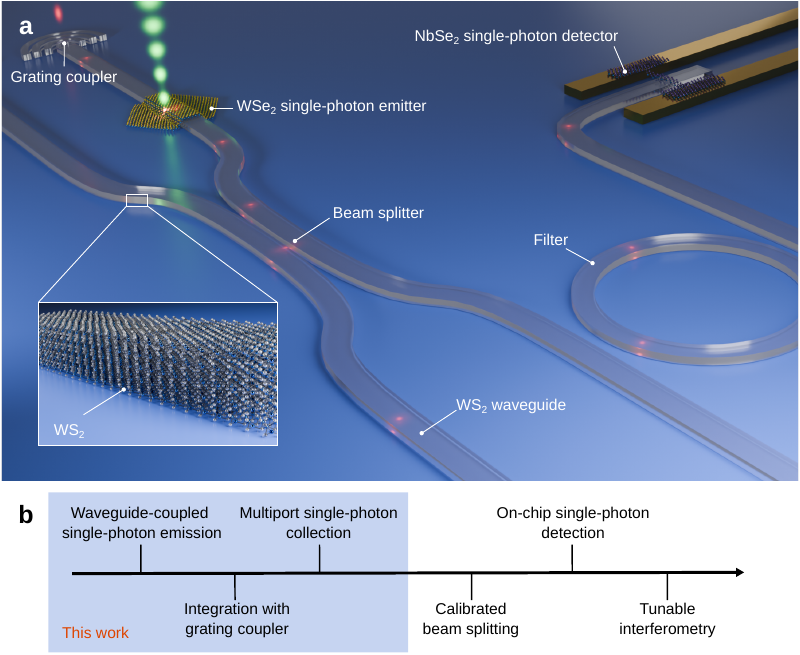}
    \caption{\rev{Future prospects of a} TMD-based photonic integrated circuit for scalable quantum information processing. \rev{(a) Illustration of a potential future implementation of a TMD-based integrated quantum photonic circuit.} Distinct optical, electronic, and mechanical properties across different TMDs allow each material to serve a specialized function in an integrated quantum photonic architecture: mono- or bilayer WSe$_2$ as single-photon emitters, nanostructured multilayer WS$_2$ as high-index and low-loss photonic integrated circuits incorporating couplers, filters, beam splitters, and interferometers, and superconducting NbSe$_2$ nanowires as fast and efficient single-photon detectors. \rev{(b) Development timeline of TMD-based integrated quantum photonics, indicating completed milestones (shaded blue area) and required future demonstrations.}
    }
    \label{fig:concept}
\end{SCfigure*}

In this work, we \rev{address the coupling problem by employing a high-index, multimode WS$_2$ waveguide}. Bilayer WSe$_2$ quantum emitters are integrated on \rev{slab} WS$_2$ waveguides, enabling efficient coupling in an all-vdW emitter–waveguide system \rev{by retaining a substantial overlap between an arbitrarily oriented dipole and TM or hybrid field modes}. \rev{Consistently with other reports \cite{paralikisTunableLowNoiseWSe22025, wyborskiTriggeredGenerationIndistinguishable2025, kleinSiteselectivelyGeneratedPhoton2019}, we observe that using hBN as a spacer between the two TMD materials is beneficial, limiting fracturing during the transfer of 2L-WSe$_2$ and increasing the emitter brightness.} 

We verify the quantum nature of a bilayer WSe$_2$ emitter by first using confocal detection, obtaining $g^{(2)}(0) = 0.043\pm0.027$. We then repeat the measurement by collecting the signal via integrated grating couplers, retaining \rev{an error-limited} $g^{(2)}(0) = 0.003^{+0.030}_{-0.003}$---at least fifty times lower than earlier reports \cite{errando-herranzResonanceFluorescenceWaveguideCoupled2021, tonndorfOnChipWaveguideCoupling2017, liIntegrationHBNQuantum2021, gerardTopdownIntegrationHBN2023}. Taking further advantage of the dual-ended waveguide, we \rev{demonstrate multiport antibunching by directly cross-correlating the two outputs of the waveguide, thereby making the beam splitter in the Hanbury Brown-Twiss (HBT) setup redundant}. High single-photon purity is preserved in this configuration ($g^{(2)}(0) = 0.076\pm0.023$)\rev{, making it the first two-port cross-correlation of a vdW single-photon emitter to the best of our knowledge}. 
Moreover, we \rev{report} an out-coupled single-photon rate at the first lens of approximately \rev{357}~kHz (\rev{259}~kHz) from the first (second) grating coupler under excitation \rev{of a higher-level discrete state}. Under the assumption of an out-coupling efficiency of \rev{21.2\%}, the upper limit obtained via numerical simulations, this translates to a \rev{minimum} waveguide-coupled rate of \rev{1.68}~MHz (\rev{1.22}~MHz). \rev{Finite-difference time-domain models attribute the observed increase in waveguide-coupled single-photon rates to the use of a multimode waveguide configuration. In addition, \rev{discrete-state} excitation enables higher single-photon purity and brighter photoluminescence by reducing the spectral background.}
\rev{These results establish a TMD platform delivering waveguide-coupled single photons with high purity and brightness,} paving the way toward \rev{hybrid}, vdW-centred quantum photonic integrated circuits.

\section{Our vision: a versatile van der Waals platform for quantum photonics}

Figure \ref{fig:concept}\rev{a} illustrates a concept for a \rev{future} photonic integrated circuit made primarily of TMD materials with complementary optical and electrical functionalities \cite{manzeli2DTransitionMetal2017, munkhbatOpticalConstantsSeveral2022}. Within this platform, strain-engineered TMDs, such as WSe$_2$, serve as on-demand quantum light sources. 
Single-photon emission has been observed in both monolayer and bilayer WSe$_2$, with bilayers offering reduced background photoluminescence while maintaining high brightness \cite{desaiStrainInducedIndirectDirect2014, kumarStrainInducedSpatialSpectral2015, piccininiHighpurityStableSinglephoton2025a}. In this concept, local strain is applied by transferring WSe$_2$ onto slab waveguides, where edge-induced wrinkles create quantum emitters, \rev{as earlier reports have shown} \cite{peyskensIntegrationSinglePhoton2019, errando-herranzResonanceFluorescenceWaveguideCoupled2021}.

Photonic circuits are realised with nanostructured multilayer TMDs, such as WS$_2$, which can be patterned with established nanofabrication techniques \cite{munkhbatNanostructuredTransitionMetal2023, munkhbatTransitionMetalDichalcogenide2020} and offer highly confined, low-loss propagation from 750 nm up to the telecom band, as required by miniaturised integrated photonics \cite{vyshnevyyVanWaalsMaterials2023}. Other photonic components, such as couplers, cavities, and spectral filters, can also be implemented with the same material \cite{munkhbatNanostructuredTransitionMetal2023, lingDeeplySubwavelengthIntegrated2023}. Finally, vdW superconductors, such as NbSe$_2$, can be patterned into integrated superconducting-nanowire single-photon detectors (SNSPDs) \cite{metuhSinglePhotonDetectionSuperconducting2025, zuglianiSinglephotonDetectionFewlayer2026}, whose detection rates, efficiency, and timing jitter excel among single-photon detector classes. 

\rev{The device presented in this work features waveguide-coupled single-photon emission in a WS$_2$/WSe$_2$ architecture, which is collected via grating coupling. By measuring the second-order cross-correlation between two output ports, we estimate the waveguide-coupled emission split along the two waveguide directions and discuss the coupling mechanisms in this system. In Figure \ref{fig:concept}b, we propose exemplary future milestones that will further validate the TMD-based platform for integrated quantum photonics, including calibrated beam splitting, single-photon detection, and interferometry.}

\rev{Despite the rapid progress in wafer-scale growth of TMDs \cite{zhangMetalOrganicChemical2025}---which plays a critical role in scaling the platform to industrial production---}such a \rev{device} can be \rev{fabricated} without the need for physical or chemical vapour deposition techniques \rev{and is therefore unconstrained by material requirements for hybrid integration}. The lack of surface dangling bonds in 2D materials results in an effective integration approach \cite{turunenQuantumPhotonicsLayered2022}, where vdW films can be  transferred on any photonic circuit with pre-patterned devices for generating \cite{blauthCouplingSinglePhotons2018, tonndorfOnChipWaveguideCoupling2017, peyskensIntegrationSinglePhoton2019}, guiding \cite{vyshnevyyVanWaalsMaterials2023, lingDeeplySubwavelengthIntegrated2023}, and detecting \cite{metuhSinglePhotonDetectionSuperconducting2025, zuglianiSinglephotonDetectionFewlayer2026} single-photon states, thus making these materials compatible with existing circuit platforms.
\rev{For instance, the integration of a single-photon detector--realised with NbSe$_2$ superconducting nanowires in Fig. \ref{fig:concept}--could be incorporated into the process flow as an intermediate step between transferring (or growing) the WS$_2$ film and patterning the circuit. Following previous demonstrations with other superconductors \cite{sprengersWaveguideSuperconductingSinglephoton2011}, NbSe$_2$ flakes can first be transferred onto unpatterned WS$_2$; after defining the contact pads, a nanowire pattern can then be fabricated via lithography and etching, followed by mask removal and completion of circuit fabrication. Transfer of the emitter-hosting material (such as WSe$_2$) would occur last, as no further top-down processing follows thereafter.}

\section{Van der Waals photonic devices: waveguides and grating couplers}

\begin{figure*}[t!]
    \centering
    \includegraphics[width=0.9\linewidth]{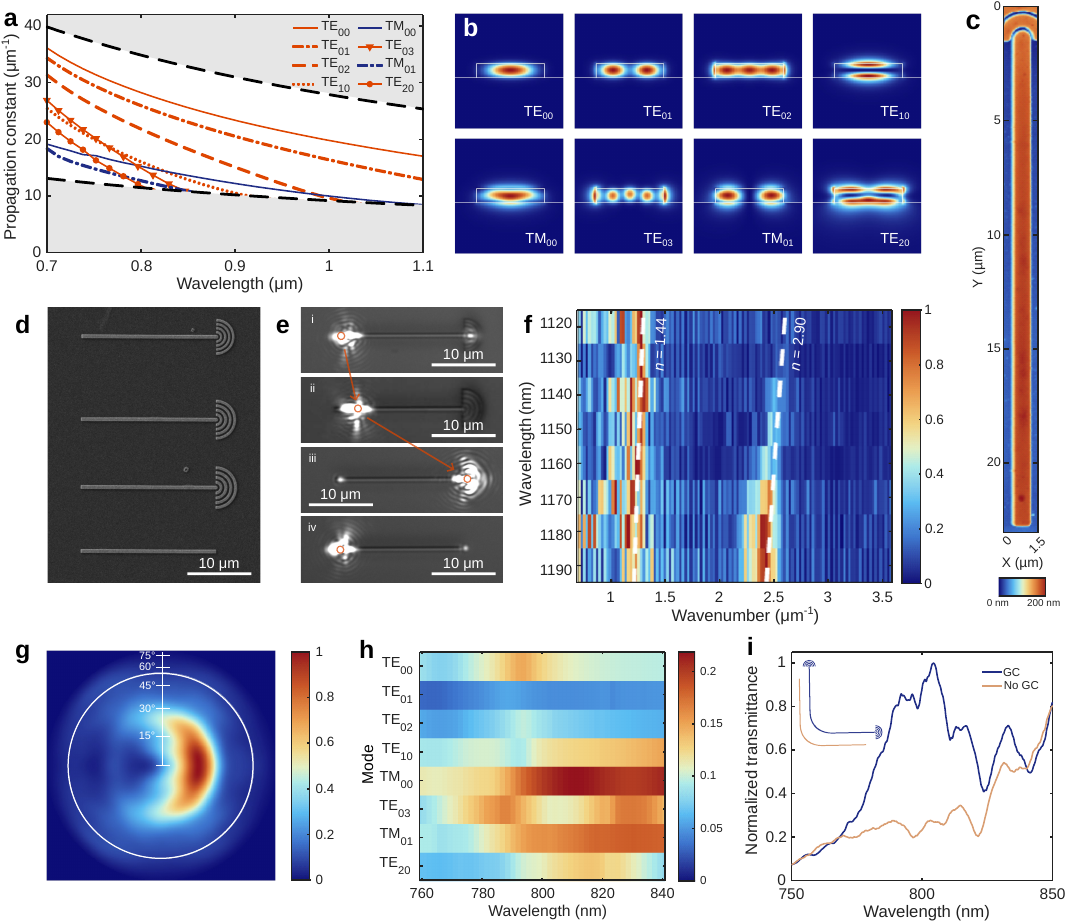}
    \caption{\textbf{Nanostructured WS$_2$ for on-chip waveguides.} (a) Dispersion relation of a $500 \times 150$~nm WS$_2$ slab waveguide on a SiO$_2$ substrate, showing all propagating modes with a cut-off wavelength above $\lambda = 800$~nm. The dashed black lines mark the propagation constant in bulk WS$_2$ and SiO$_2$. (b) Electric field intensity profile at $\lambda = 802$~nm for each propagating mode shown in (a); the waveguide and substrate cross-sections are outlined in white. (c) Atomic force microscopy trace of the $500 \times 175$~nm waveguide used for near-field microscopy, showing a slab waveguide with minimal surface roughness. (d) Scanning electron microscopy image of short $500 \times 150$~nm waveguides with variable grating coupler design or no grating coupler (bottom waveguide). (e)  Optical microscopy image of two of the waveguides shown in (d) under illumination with a broadband source; the empty orange circle indicates the beam spot. Light can be coupled in the waveguide by edge scattering (i, iv) or grating coupling (iii) and observed on the opposite end; no light is coupled by irradiating the surface of the waveguide (ii). (f) Dispersion relation of the TE modes measured with near-field microscopy (sSNOM) over the waveguide in (c) at decreasing photon wavelengths, from 1120 to 1190~nm. The two dashed lines matching the spectral features correspond to an effective refractive index of 2.90 (propagating mode) and 1.44 (SiO$_2$ substrate mode). (g) Far-field distribution of the upward-coupled fundamental TE mode at $\lambda = 800$~nm from the proposed grating coupler design, obtained via FDTD modelling; the light within the white circle can be collected with an objective with $\text{NA} = 0.81$. (h) FDTD modelling of the collection efficiency spectra for the propagating modes with the \rev{fabricated grating coupler geometry}.  (i) Comparison of the total transmittance between grating coupling (blue) and side-scattering coupling (yellow) through a 60~\textmu m-long WS$_2$ waveguide. The spectra are ten point-averaged and normalized by the grating-coupled transmittance value at 800~nm. The inset shows an illustration of the waveguides characterised in the measurement.}
    \label{fig:passive}
\end{figure*}

We begin by establishing multilayer WS$_2$-based optical waveguides and grating couplers as building blocks for integrated photonics. The proposed geometry consists of multimode slab waveguides ($w = 500$~nm, $h = 150$~nm), designed for large mode overlap with dipole-like emitters located near the waveguide. Figures \ref{fig:passive}a,b show the dispersion relation of the propagating modes and their electric field profile at 802~nm, respectively. Despite separating the modes into transverse-electric (TE) and transverse-magnetic (TM) depending on their primary polarisation, it is worth noting that at 802~nm most higher-order modes are hybrid, and therefore have a variable polarisation ratio. 
The waveguide fabrication process begins by mechanical exfoliation of WS$_2$ flakes. When suitable flakes at the desired thickness \rev{(measured as described in Supplementary Note 1)} are found, they are transferred onto Si chips with a 2~\textmu m SiO$_2$ layer and pre-patterned alignment markers. The designed structures are then fabricated by transferring an electron-beam resist pattern onto the flake via reactive-ion etching (Supplementary Note 1). 
The resulting waveguides are characterised by a smooth surface (given by the pristine WS$_2$ flake) and limited sidewall roughness or etching residues, as reported in atomic force microscopy (AFM) and scanning electron microscopy (SEM) images in Fig. \ref{fig:passive}c,d. 

Initially, we illustrate the possible coupling mechanisms with a set of three short (21~\textmu m) waveguides with a single-sided grating coupler, whose design is described later in this section (Fig. \ref{fig:passive}d). As shown in Fig. \ref{fig:passive}e, a broadband objective-focused beam can be launched into the waveguide by scattering through the ungrated end, couple out of a grating coupler (as revealed by the smaller bright spot), and be collected by the same objective (i). If the beam spot is moved slightly, no light scatters through the waveguide (ii). Light can also be injected via the grated end (iii), and emission can be observed because of the up-scattered light from the opposite end. Finally, scattering from the waveguide ends enables coupling to and from a high-numerical aperture (NA) objective even without gratings (iv), despite the reduced efficiency. 

To examine propagation (shown in Fig. \ref{fig:passive}c), we perform scattering-type scanning near-field optical microscopy (sSNOM) between 1120--1190~nm on a 20~\textmu m-long waveguide  (see Methods and Supplementary Fig. 3).
For each wavelength, we extracted the momenta of the propagating waves \cite{cassesQuantitativeNearfieldCharacterization2022, cassesFullQuantitativeNearField2024}. and plotted them against the excitation wavelength to reconstruct the waveguide-mode dispersion (Fig. \ref{fig:passive}f). This revealed a mode with wavenumber $q = 2.45~$\textmu$ \text{m}^{-1}$ at $\lambda = 1190$~nm, corresponding to an effective index $n_\text{exp} = 2.90$.
By comparison, the simulated index of TE$_{00}$ in a WS$_2$ slab waveguide with $500 \times 175$~nm in cross-section and 40~nm of over-etch on SiO$_2$ is $n_\text{num} = 2.99$, in good agreement with the \nobreak{sSNOM} data. A second prominent feature corresponds to $n_\text{sub} = 1.44$, a substrate mode in the SiO$_2$ layer.

When probing waveguide-coupled emitters at low temperature, using a single optical path for excitation and collection via a high-NA objective is convenient to remove the need for separate fibres at each port.
However, collection via an objective is less efficient than optical fibre coupling, as a fibre can be independently aligned to the phase-matched out-coupled beam \cite{chengGratingCouplersSilicon2020}. In addition, typical grating designs are optimised for fibre-core dimensions and include a tapered waveguide to adiabatically alter the field confinement \cite{chengGratingCouplersSilicon2020}. We therefore adopted a compact grating coupler geometry suited to vertical microscope operation, consisting of concentric half-rings \cite{zhuUltracompactSiliconNitride2017}. The higher effective index in WS$_2$ compared with the referenced work \cite{zhuUltracompactSiliconNitride2017}, which proposes a silicon-nitride coupler, causes the main deviation from the original design. The single-step etching process was maintained, thus fixing the grating depth to the waveguide height.

To maximise extraction of waveguide-coupled emission, we optimised the coupling efficiency from a \rev{TE$_{00}$} guided mode to an objective with $\mathrm{NA} = 0.81$ by sweeping the grating parameters in finite-difference time-domain (FDTD) simulations (Methods).
The optimised coupler comprises four half-rings with a 780~nm pitch and a 55\% duty cycle \rev{for optimal extraction of the TE$_{00}$ mode at 800~nm (Supplementary Fig. 4)} within the $5\rev{4}.1^\circ$ half-angle cone corresponding to the objective NA. 
The TM$_{00}$ mode\rev{and other higher-order modes with a TM component} exhibit a similar pattern, which is however nearly orthogonal in polarisation to TE$_{00}$ (Supplementary Fig. 5).  
\rev{To account for fabrication variations, AFM and SEM scans of the grating couplers were used to determine the real pitch, duty cycle, and etch depth, which were used to correct the geometry in the simulation model.} Extraction-efficiency spectra \rev{of the corrected model} (Fig. \ref{fig:passive}h) give an average of  \rev{12.2\%} at 802~nm across all propagating modes, with a maximum of \rev{21.2\%} for TM$_{00}$. Efficiency is mainly limited by the high index contrast between WS$_2$ and air, remediable by varying the grating depth (thus requiring a two-step etching process) \cite{chenFabricationTolerantWaveguideChirped2008, chengGratingCouplersSilicon2020}, and the scattering toward the substrate, which could be decreased with a suspended design or a reflective layer \cite{zhuUltracompactSiliconNitride2017}.

Light injection through WS$_2$ waveguides was characterised by comparing the spectral transmittance of grated and ungrated 60~\textmu m-long waveguides (Fig. \ref{fig:passive}i) when a focused beam entered one end and the out-coupled spectrum was measured at the opposite end. Notably, we observe a maximum around 800~nm and consistently higher transmittance between 750--850~nm with the grated structure. Extended-range spectra are provided in Supplementary Fig. 6.

\section{Integration with quantum emitters}

\begin{figure*}[t!]
    \centering
    \includegraphics[width=0.9\linewidth]{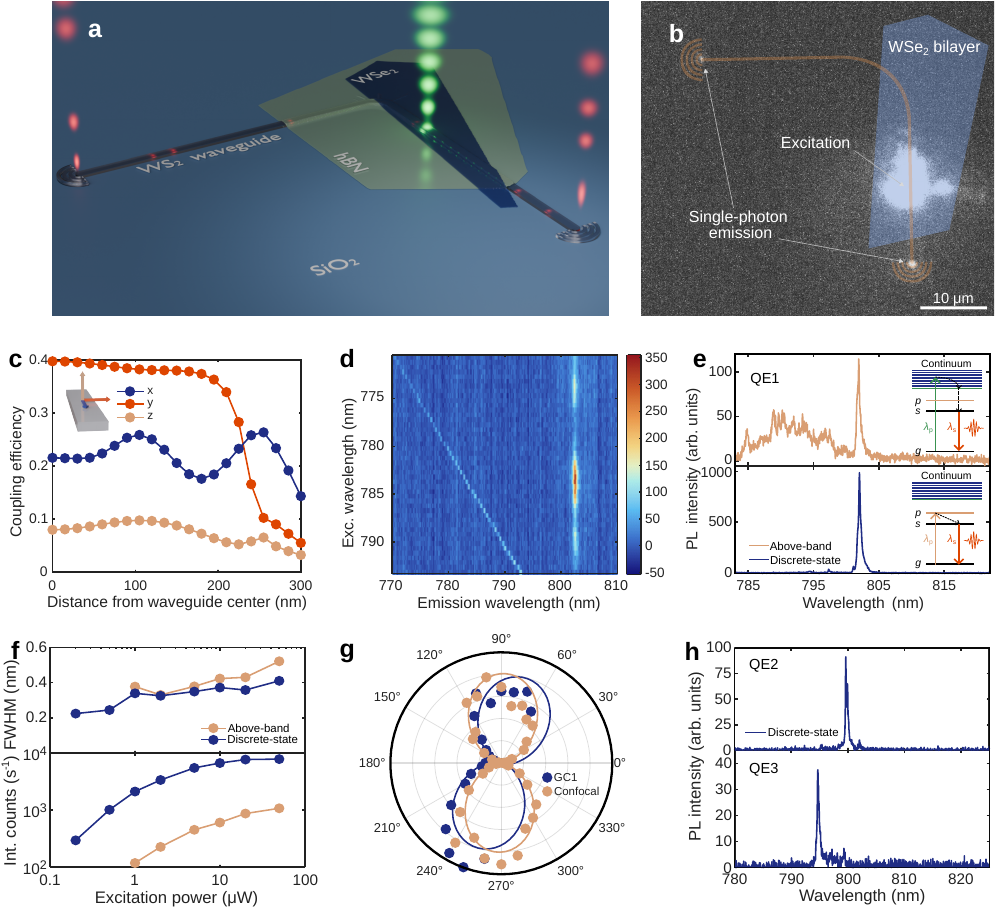}
    \caption{\textbf{Integration of WS$_2$ waveguides with hBN/WSe$_2$ heterostructures for on-chip coupling of single photons.} 
    (a) Artistic illustration of the characterised device, consisting of a 60~\textmu m-long WS$_2$ waveguide with a $90^\circ$ bend and grating couplers, with a thin ($<5$~nm) hBN layer and a WSe$_2$ bilayer transferred on top. By optically exciting a quantum emitter in the WSe$_2$ bilayer, single photons couple to waveguide modes and are collected via the grating couplers. 
    (b) Photoluminescence emitted by a laser-excited quantum emitter on bilayer WSe$_2$ emitter; the waveguide and a portion of the bilayer are overlapped in the image for reference. Photoluminescence is observed on both the excitation spot and the two grating couplers. 
    (c) Simulated \rev{emitter-waveguide} coupling efficiency \rev{$\beta_+$} between an electric dipole 5~nm above the waveguide and a \rev{forward} propagating mode as a function of the distance from the vertical symmetry axis of the waveguide cross-section; the coupling is calculated with the dipole oriented along each main axis. 
    (d) Photoluminescence excitation spectral map of a quantum emitter at $\lambda = 802$~nm, measured from GC1; the emitter can be excited via \rev{discrete-state} excitation, showing peak brightness with an excitation laser at $\lambda_p = 783$~nm. The partially-filtered laser line is intentionally shown. 
    (e) Emission spectra from GC1 obtained via above-band and \rev{discrete-state} excitation of the quantum emitter\rev{, with the insets showing the energy diagram under the two excitation schemes;} \rev{discrete-state} excitation suppresses the majority of the background emission and increases the emitter brightness. 
    (f) Power-dependent photoluminescence (integrated counts per second at the spectrometer, logarithmic scale) and spectral linewidth of the quantum emitter at GC1 under the two excitation schemes.
    (g) Polarisation-dependent photoluminescence at GC1 or confocal to the excitation laser under \rev{discrete-state} excitation, fitted with a sinusoidal curve.
    (h) Additional examples of spectral lines coupled through the waveguide and collected from a grating coupler.}
    \label{fig:integration}
\end{figure*}

After inspecting the fabricated structures, we integrated quantum emitters on the 60~\textmu m-long waveguide characterised in Fig. \ref{fig:passive}i by transferring a thin hexagonal boron nitride (hBN) layer onto part of the waveguide, followed by a WSe$_2$ bilayer\rev{, whose thickness is identified from its room-temperature photoluminescence emission (Supplementary Note 3)} \cite{tonndorfPhotoluminescenceEmissionRaman2013}.
By having a thin layer of hBN between WS$_2$ and WSe$_2$, the bilayer can be transferred undamaged, while maintaining proximity between the two materials to facilitate the emitter coupling into the waveguide. \rev{By contrast, 2L-WSe$_2$ flakes directly transferred onto WS$_2$ waveguides resulted in multiple fractures in the bilayer. In addition, WSe$_2$ emitters on a hBN spacer were consistently brighter than their bare counterparts. Further details on the role of hBN are provided in Supplementary Note 6.}

\rev{After the transfers,} the sample was cooled to $T=4$~K. Emitters were located by scanning a laser along the waveguide and monitoring confocal photoluminescence; once a spectral line \rev{was found}, the collection spot was repositioned to either grating to verify the emitter-waveguide coupling (Fig. \ref{fig:integration}a). Waveguide-coupled photoluminescence can also be imaged with a camera, as shown in Fig. \ref{fig:integration}b, where the laser is positioned on top of the quantum emitter characterised in Fig. \ref{fig:integration}d--g. Additional photoluminescence images are provided in Supplementary Fig. 9.

To assess how position and orientation of the emitter---modelled as a dipole source \cite{wangNearunityEfficiencyRidge2022}---affect coupling efficiency, we performed FDTD simulations to calculate the coupling between the dipole and each propagating mode. The results show that overall TM and hybrid modes couple more strongly to the dipole because of a larger field overlap (Supplementary Note 5). Furthermore, the efficiency varies markedly with dipole orientation; Figure \ref{fig:integration}c reports the total coupling efficiency to any guided mode for each dipole orientation axis. Due to the hybrid polarisation of some propagating modes, coupling is possible not only for an in-plane dipole orthogonal to the waveguide axis ($y$), but also for a dipole oriented along the waveguide ($x$) and out of plane ($z$), albeit with reduced efficiency.
Since dipole orientation in strained WSe$_2$ flakes tends to follow the wrinkle direction \cite{paralikisTailoringPolarizationWSe22024}, transferring a flake onto a slab waveguide might induce emitters that are oriented along the waveguide edge. The emitter-waveguide coupling has also been modelled with a dipole along the side of the waveguide rather than above it (Supplementary Note 5). 

We now focus on a spectral line at $\lambda = 802$~nm, whose emitter (QE1) lies on the excitation spot shown in Fig. \ref{fig:integration}b, and compare two excitation schemes: above-band excitation (ABE) at 650~nm and \rev{discrete-state} excitation (\rev{DSE}), obtained by \rev{scanning the pump energy well between the bandgap of 2L-WSe$_2$ (747~nm) \cite{terronesBilayersTransitionMetal2014} and the emission line (802~nm). We target the excited state with the higher-energy discrete level enabling the brightest photoluminescence, while maintaining at least 10~nm of spectral separation with the emitter,} thereby still allowing efficient spectral filtering of the pump. \rev{As shown in the spectral map of Fig. \ref{fig:integration}d, this corresponds to $\lambda_p = 783$~nm for QE1 (38~meV above the ground state emission), a detuning consistent with previous studies of higher-order electronic shell excitation in WSe$_2$  \cite{mitryakhinRabiOscillationsMonolayer2025}. At this excitation energy,} considerably brighter emission is measured both with a confocal collection spot and at the grating couplers. 
Two spectra of QE1 collected at the upper grating coupler (GC1) are compared in Fig. \ref{fig:integration}e under ABE and \rev{DSE} at near-saturation pump power\rev{, where illustrations of the energy diagrams for the two schemes are presented as insets}. Brighter emission is achieved under \rev{DSE}, while simultaneously reducing the spectral background from higher-energy states. \rev{The integrated counts of QE1 under \rev{DSE} increased by roughly one order of magnitude at equal pump power, while the emitter linewidth is comparable to ABE, if slightly narrower (Fig. \ref{fig:integration}f and Supplementary Fig. 10). We attribute the brighter photoluminescence under \rev{DSE} to the decrease of non-localised background emission and of non-radiative relaxation channels by pumping well below the bandgap energy \cite{terronesBilayersTransitionMetal2014, mitryakhinRabiOscillationsMonolayer2025}, and to a shorter decay time of the emitter, from $\tau_\text{ABE} = (55.0\pm1.1)$~ns to $\tau_\text{\rev{DSE}} = (31.1\pm0.4)$~ns (Supplementary Fig. 11).} \rev{Power-dependent photoluminescence fitted to $I = I_\text{sat} P^\alpha/(P+P_\text{sat})$ yielded $\alpha = 0.91\pm 0.07$ under DSE, a value consistent with a basic excitonic transition.}
Both excitation schemes yield brighter waveguide-coupled photoluminescence of TMD single-photon emitters than previously reported \cite{errando-herranzResonanceFluorescenceWaveguideCoupled2021, peyskensIntegrationSinglePhoton2019}, which we partly attribute to a larger field overlap between the emitter and the guided modes \cite{michaelisdevasconcellosSinglePhotonEmittersLayered2022}. \rev{Notably}, bilayer WSe$_2$ transferred on WS$_2$ waveguides without a bottom hBN flake coupled more weakly (Supplementary Note 6), suggesting a role of hBN in affecting the WSe$_2$ charge environment \cite{paralikisTunableLowNoiseWSe22025}.

\begin{figure*}[t!]
    \centering
    \includegraphics[width=0.9\linewidth]{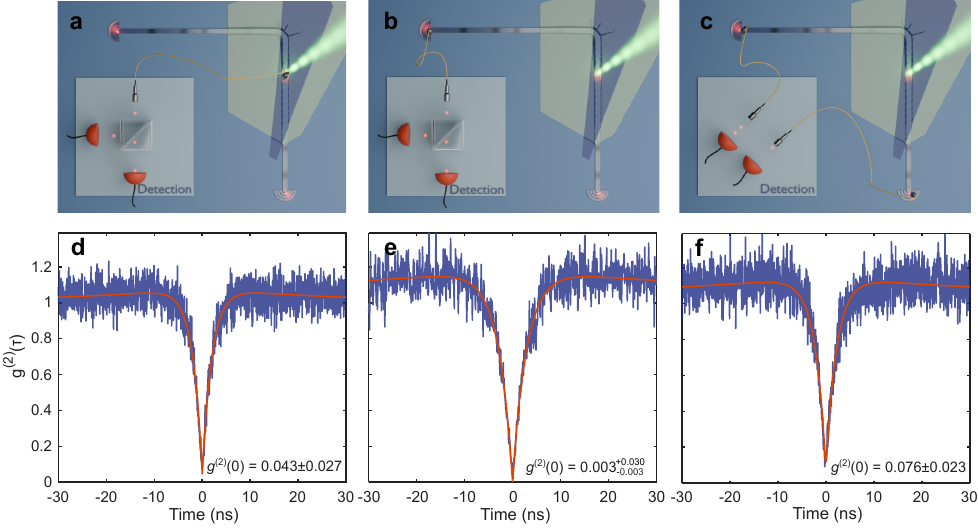}
    \caption{\textbf{On-chip waveguide-coupled single-photon purity.} (a-c)
    Simplified illustration of the Hanbury Brown and Twiss (HBT) experimental setup realised with (a) the collection spot confocal to the excitation source and an external beam splitter before the detectors; (b) the collection spot at one grating coupler with an external beam splitter before the detectors; (c) a collection spot at each grating coupler, so that the HBT effect can be observed without an external beam splitter. (d-f) Continuous-wave second-order correlation traces obtained with the respective setup; the traces are plotted with a three-point moving average, and show signatures of photoluminescence blinking. All configuration exhibit high single-photon purity at zero time delay.}
    \label{fig:hbt}
\end{figure*}

Having established the efficacy of \rev{DSE}, we evaluate additional spectral properties of waveguide-coupled WSe$_2$ emitters.
A photoluminescence spectrogram (Supplementary Fig. 12) shows limited spectral wandering, which could be further reduced by stabilising the charge environment \cite{paralikisTunableLowNoiseWSe22025}. 
In addition, polarisation-resolved measurements confirm the polarised nature of the emitter, and polarisation is preserved even for waveguide-coupled emission (Fig. \ref{fig:integration}g and Supplementary Fig. 13). The degree of linear polarisation (DOLP), calculated as $(I_\text{max}-I_\text{min})/(I_\text{max}+I_\text{min})$, at GC1 and confocal to the emitter gives $(84\pm18)\%$ and $(98^{+2}_{-8})\%$, respectively, similarly to other strain-engineered WSe$_2$ emitters \cite{paralikisTailoringPolarizationWSe22024}.
The polarisation aligns with the waveguide in the confocal case, implying that the dipole has a component parallel to the waveguide. However, an out-of-plane component cannot be excluded, since its emission would be largely inaccessible under confocal collection yet could still couple to a waveguide mode. Both dipole orientations couple most efficiently with TM or hybrid modes. In addition, since the out-coupled polarisation at the grating couplers matches the simulated field distribution of \rev{modes with a large TM component} (Supplementary Fig. 5), we conclude that a TM-polarised waveguide mode is the main contributor to the measured signal.

Finally, to further demonstrate the fabrication of waveguide-coupled quantum emitters with this method, we report additional bright lines between 795 and 800~nm from other emitters (Fig. \ref{fig:integration}h) under \rev{DSE} and collection at a grating coupler, confirming that WS$_2$ waveguides constitute a robust platform for on-chip coupling of vdW emitters. 

\section{On-chip second-order correlation measurement}

To verify the single-photon nature of QE1, we evaluated the second-order autocorrelation function $g^{(2)}(\tau)$ under different HBT configurations.
First, single photons are collected confocally from the excitation spot (Fig. \ref{fig:hbt}a). The signal is then split into two arms with a beam splitter and correlated after detection with an SNSPD, producing the autocorrelation trace shown in Fig. \ref{fig:hbt}d.
The fitted curve yields a zero-delay second-order correlation of $g^{(2)}(0) = 0.043\pm0.027$, demonstrating the high single-photon purity of QE1.

The measurement is repeated by moving the collection spot to GC1 (Fig. \ref{fig:hbt}b), thereby correlating the waveguide-coupled emission. With this dataset, the purity of the single-photon state ($g^{(2)}(0) = 0.003^{+0.030}_{-0.003}$) increases further, approaching the unity limit. We attribute the increased purity from the confocal configuration to a slightly lower excitation power and to more efficient waveguide coupling of the emitter compared with background emission.

In the final configuration, the beam splitter in the HBT setup is removed, and a second collection spot is implemented and aligned with the other grating coupler, GC2 (Fig. \ref{fig:hbt}c), \rev{thereby enabling two-port cross-correlation of the single-photon emitter}. In addition to simplifying the experimental setup, this improves the robustness of the second-order autocorrelation measurement, which is obtained by correlating photons at two separate waveguide ports, rather than halving the photon stream from a single port. The resulting measurement confirms the preserved purity of the quantum emitter ($g^{(2)}(0) = 0.076\pm0.023$) and provides, to our knowledge, the first demonstration of an HBT measurement of vdW quantum emitters \rev{correlating two waveguide ports}. The whole second-order correlation traces and raw $g^{(2)}(0)$ values are presented in Supplementary Note 7.

We conclude by discussing \rev{single-photon rates} and brightness. The single-photon count rate at saturation measured by the SNSPD during the HBT measurement yielded \rev{20.0} and \rev{14.5}~kHz for GC1 and GC2, respectively, \rev{corresponding to a split of 0.58:0.42.} Accounting for \rev{transmission and detection} losses in the \rev{experimental} setup, we \rev{obtained} a single-photon rate at the first lens of \rev{357 and 259}~kHz for the respective grating coupler. Since the \rev{modelled grating} efficiency \rev{of the fabricated device at 802~nm} is \rev{$\leq21.2\%$}, as shown in Fig. \ref{fig:passive}h, we \rev{determine a lower bound on the} waveguide-coupled single-photon rate of \rev{$\geq1.69$}~MHz and \rev{$\geq1.22$}~MHz along the two directions. We note that this is a conservative estimation, as fabrication imperfections and multimode coupling render the simulated extraction efficiency value an upper bound. In comparison, the confocal count rate at the first lens is evaluated at \rev{1.79}~MHz. Given an average re-excitation time of \rev{1.84}~ns (from the \rev{continuous-wave antibunching rise} time), the total \rev{waveguide brightness, which includes losses due to propagation, emitter-waveguide coupling, and internal quantum efficiency,} is \rev{bounded as $\geq0.65\%$}.
Details about the single-photon rates and efficiency figures are provided in Supplementary Note 8. 
Further improvements could be \rev{achieved} by increasing spectral stability via electrical biasing. In addition, emitter brightness and indistinguishability could be enhanced by integration with high-$Q$ optical cavities \cite{iffPurcellEnhancedSinglePhoton2021, binkowskiHighPurcellEnhancement2025, kimPhotonicCrystalCavities2018} and advanced excitation schemes \cite{piccininiHighpurityStableSinglephoton2025a, vannucciSinglephotonEmittersWSe2024}.

\section{Conclusions}

In this work, we demonstrated integrated quantum emitters built entirely on a van der Waals platform. The integration of bilayer WSe$_2$ quantum emitters with multimode WS$_2$ waveguides enabled efficient on-chip coupling of high-purity, polarised single photons. Multiple single-photon emitters \rev{within a single WSe$_2$ flake} were shown to couple efficiently to WS$_2$ waveguides, highlighting the robustness of the platform. Moreover, the bidirectional coupling of the WS$_2$ waveguide was exploited to establish a \rev{two-port} second-order correlation experiment. The MHz-level waveguide-coupled single-photon rates indicate efficient emitter-waveguide coupling and underscore the potential of the vdW platform for bright integrated sources. Looking ahead, \rev{systematic studies of transfer yield---determining coupled emitter density, spectral distribution, and coupling efficiency per emitter---would further validate the platform. Furthermore,} extending this platform with single-photon detectors, phase modulators, and interferometric components would enable a promising route to vdW quantum photonic integrated circuits, paving the way for scalable quantum information processing based entirely on nanoengineered two-dimensional materials.

%TC:ignore
\section*{Methods} % Up to 3000 words

\subsection{Numerical modelling}

\textbf{Eigenmode modelling.} Modelling of the propagating modes and the dispersion relation were carried out with a finite-difference eigenmode solver in Ansys Lumerical. For the calculations, a dispersive complex refractive index was fitted for each material; for WS$_2$, an anisotropic index was implemented \cite{munkhbatOpticalConstantsSeveral2022}. 

\textbf{Grating coupler simulations.} The design of the grating coupler and the calculations of waveguide-objective coupling, far-field maps, and dipole-waveguide coupling were simulated with the FDTD solver in Ansys Lumerical. The initial structure of the grating coupler was adopted by another work \cite{zhuUltracompactSiliconNitride2017} and optimised by sweeping ring period, duty cycle (i.e., the ratio between the ring width and the ring period), and number of rings. 
The out-coupling efficiency, which is the product of the upward-scattered transmittance from the grating with the fraction of the far field within the objective aperture, was calculated by calculating the transmittance from each propagating mode to a field monitor above the grating coupler; the transmittance was then integrated within the light cone of an objective with $\text{NA} = 0.81$, corresponding to a half-angle of about $53.1$°.

\textbf{Dipole-waveguide coupling simulations.} Dipole-waveguide simulations were modelled in Ansys Lumerical by calculating the projection of each propagating mode in the transmittance from a dipole source into \rev{one port of} the slab waveguide. Due to the variable dipole power emitted by the source (which is highly dependent on the dipole position relative to the waveguide), the coupling efficiency was normalized by the Purcell factor at each position, so that the coupling relative to the real emitted power can be evaluated \cite{wangNearunityEfficiencyRidge2022}. Unless otherwise stated, the dipole was positioned 5~nm above the waveguide to mimic the hBN thickness.
The total coupling efficiency is calculated by summing the contribution of all eight propagating modes for each dipole axis orientation.

\subsection{Nanofabrication}

\textbf{Substrate preparation.} The substrates are prepared by thermally growing 2~\textmu m of SiO$_2$ on silicon and patterning alignment markers with a negative UV resist (AZ nLOF 2020) and electron-beam evaporation of 5~nm of titanium, followed by 50~nm of gold. The metal is lifted off in an ultrasonic bath of MICROPOSIT Remover 1165 and rinsed in isopropyl alcohol and water. 

\textbf{Exfoliation and transferring.} WS$_2$, hBN, and WSe$_2$ (HQ Graphene) are exfoliated using the same stamping technique \cite{castellanos-gomezDeterministicTransferTwodimensional2014}. Semiconductor wafer tape (Nitto SWT 20+R) is used to initially exfoliate and thin down large flakes from a bulk crystal. Polydimethylsiloxane (PDMS) stamps on glass slides are then used to exfoliate the crystal into smaller flakes, which are selected by thickness and homogeneity.
The thickness of WS$_2$ flakes was measured with the transfer matrix method by fitting reflectivity spectra from confocal illumination of the flake on the PDMS stamp with a broadband source (Supplementary Note 1). Thin flakes of hBN were selected by their optical contrast under an optical microscope, and bilayer WSe$_2$ was identified by room-temperature photoluminescence with a blue LED (Supplementary Note 3).

WS$_2$ flakes are positioned and transferred onto the substrate with a transfer station (HQ Graphene Systems HQ2D MAN), where the substrate is heated to 70°C to overcome the adhesion of the flake to PDMS upon contact. Three baths of acetone, isopropyl alcohol, and deionised water are used to clean the sample from contaminants at the end of the transfer process. After transferring the flakes, the fit was confirmed with a stylus profilometer (Bruker Dektak XTA).
The same method has been used to transfer hBN and WSe$_2$ after patterning the waveguides and removing the resist mask.  

\textbf{Patterning of WS$_2$ waveguides.} WS$_2$ flakes were patterned with electron-beam lithography by spin coating an adhesion promoter (Microchemicals TI Prime) and negative resist (Allresist AR-N 7520) and exposing it (Raith eLINE at 30~kV or JEOL JBX-9500FS at 100~kV). A thin aluminium layer (10~nm) was thermally evaporated on the samples processed at 100~kV before the exposure.
Aluminium was removed with AZ 726 MIF (2.38\% tetramethylammonium hydroxide) in 30~s and rinsed in water, while the resist was developed with Allresist AR 300-47 in 1 min and rinsed in water. The resist was descummed in an oxygen plasma for 30~s.
The pattern was transferred to the WS$_2$ flake with an ICP-RIE tool (SPS M/PLEX ICP--RIE) in a CHF$_3$- and Ar-based chemistry (40~sccm flow for each gas, 10~mTorr pressure, 150~W coil power and 25~W platen power), and the resist is removed with an oxygen plasma.

\subsection{Characterisation}

\textbf{Near-field measurements.} A scattering-type scanning near-field optical microscope (attocube neaSCOPE) in reflection mode was used to map a $500 \times 175$~nm WS$_2$ slab waveguide with 50~nm of SiO$_2$ overetch. The incident light from a tunable laser (Sacher Lasertechnik Group TEC-520-1180-030) is focused by a parabolic mirror on a platinum–iridium coated silicon AFM probe tip (NanoWorld ARROW-NCPt) operating in tapping mode. The excitation light impinges in the YZ plane (see Fig. \ref{fig:passive}c) at an angle of approximately 45°. The scattered light is detected with a photodiode (Newport 2053-FC). Both the excitation and the scattered light are filtered with co-polarised linear polarisers, aligned in s-polarisation. The signal recorded by the photodiode is demodulated at the third harmonic order of the tip tapping frequency \cite{raschkeAperturelessNearfieldOptical2003, knollEnhancedDielectricContrast2000}. Together with a pseudo-heterodyne detection scheme \cite{ocelicPseudoheterodyneDetectionBackgroundfree2006}, this allows retrieval of the near-field amplitude while the far-field signal is largely suppressed.

\textbf{Waveguide characterisation.} Room-temperature coupling to the waveguide was performed with a homemade confocal photoluminescence setup by using a filtered broadband source (Thorlabs OSL2IR) coupled to a single-mode fibre (Thorlabs P3-780A-FC-2), a $50\times$ objective (Nikon LU Plan Fluor, NA = 0.8), and an infrared spectrometer (Andor Kymera 328i, Andor iDus InGaAs 1.7~\textmu m ). The sample was imaged with a CMOS camera (Thorlabs CS165MU/M).

For low-temperature characterisation measurements, the sample was loaded in an optical, closed-cycle vacuum cryostat (attocube attoDRY800) with a nano-positioner and a base temperature of $T \approx 4$~K. At low temperature, the transmission measurements were carried out with the same light source, but with a visible-light spectrometer (Horiba iHR 550, Horiba Syncerity BI-NIR) and imaging camera (Tucsen Dhyana 400D). A low-temperature objective (attocube LT-APO/NIR/0.81, $\text{NA} = 0.81$) is used to focus the light on the sample surface. The spectra were post-processed with a ten-point moving average to smoothen the signal, and are normalized by dividing the traces with a reference spectrum of a reflective gold mirror.

\textbf{Photoluminescence measurements.} Bilayer WSe$_2$ is illuminated with either a 650-nm diode laser (PicoQuant LDH-D-C-650) in continuous-wave or pulsed mode for above-band excitation, a tunable (770-795~nm) CW laser (Toptica DL PRO) for continuous-wave \rev{discrete-state} excitation, and a tunable femtosecond laser (Coherent Chameleon Ultra II) for pulsed \rev{discrete-state} excitation. The laser polarisation is controlled with a half-wave plate and a linear polariser before entering the optical window of the cryostat and is focused by the same low-temperature objective. The same objective collimates the photoluminescence signal, which is filtered from the laser pump with different configurations of bandpass or longpass filters. The detection polarisation is controlled with a half-wave plate and a linear polariser before coupling to a single-mode fibre. Photoluminescence mapping was performed with the camera and the spectrometer mentioned above.

The polarisation-resolved measurements were fitted with the following function:
\begin{equation} \label{eq:dolp}
    I(\phi) = a\sin^2(\phi-\phi_0)+b.
\end{equation}
The error in the degree of linear polarisation, calculated as $a/(a+2b)$, is given as a 95\% confidence interval and is calculated by propagating the error from \eqref{eq:dolp}.

\textbf{Decay time and second-order correlation measurements.} Time-correlation measurements were carried out by detecting the signal using a superconducting nanowire single-photon detector (ID Quantique ID281). Temporal events for time-resolved photoluminescence and second-order autocorrelation measurements were correlated with a time-to-digital converter (Swabian Instruments Time Tagger Ultra).

All second-order correlations were measured with the tunable CW laser at $\lambda_p = 783$~nm. The presented raw data are processed with a three-point average. The fitting was carried out with a three-level model, which is described by the following equation \cite{rodiekExperimentalRealizationAbsolute2017}:
\begin{equation}
    g^{(2)}(\tau) = 1 - (1+a)e^{-|\tau|/\tau_1} + be^{-|\tau|/\tau_2},
\end{equation}
where $\tau_1$ and $\tau_2$ are antibunching- and bunching-related time constants, respectively. The error in the second-order correlation at zero delay is given as a 95\% confidence interval, and is calculated as a sum of the errors of $a$ and $b$. 

\textbf{Sample imaging.} Optical microscopy images were taken in confocal microscopes with various objectives and cameras. The topography of the photonic circuits and of the transferred hBN and WSe$_2$ was characterised with a scanning electron microscope (Zeiss Supra 40 VP, 5 kV, SE2 detector) and an atomic force microscope (Bruker Dimension Icon, tapping mode with Tap150Al-G tips). Samples containing bilayer WSe$_2$ flakes were imaged only at the sample end-of-life to avoid inducing defects or charging the bilayer before characterisation.

\section*{Acknowledgements}
P.M., P.W., A.P., N.G., and B.M. acknowledge support from the European Research Council (ERC-StG ``TuneTMD'', grant no. 101076437) and Villum Fonden (project no. VIL53033). The authors also acknowledge the European Research Council (ERC-CoG ``Unity'', grant no. 865230), the TICRA foundation, the Innovation Fund Denmark (QLIGHT, no. 4356-00002A), and the cleanroom facilities at the Danish National Centre for Nano Fabrication and Characterisation (DTU Nanolab). 
F.S. and N.S. are supported by the Danish National Research Foundation through NanoPhoton—Center for Nanophotonics, Grant No. DNRF147. N.S. acknowledges funding by the Novo Nordisk Foundation NERD Programme (Project QuDec NNF23OC0082957).
Finally, P.M. acknowledges Ryo Mizuta Graphics for the Optical Components Pack asset, which was partially used for the artistic illustrations in Fig. \ref{fig:hbt}.

\section*{Author contribution}

P.M. performed the numerical modelling, fabricated the waveguides, and processed and analyzed all the presented data. F.S. performed the sSNOM characterisation, supervised by N.S. A.P. and P.M. transferred the hBN/WSe$_2$ stacks on the waveguides. P.W. and P.M. performed the optical characterisation of the waveguides and the quantum emitters. B.M. supervised and coordinated the project. P.M., P.W., A.P., N.G., and B.M. wrote the manuscript with input from all co-authors.

\section*{Competing Interests}
The authors declare they have no competing interests.

\section*{References}
\bibliography{biblio}
%TC:endignore
\appendix

\end{document}